\documentclass[a4paper, USenglish, cleveref]{lipics-v2021}

\pdfoutput=1
\hideLIPIcs
\nolinenumbers

\title{Scott's Representation Theorem and the Univalent Karoubi Envelope}
\titlerunning{Scott, Karoubi, and Univalence}

\author{Arnoud {van der Leer}}{Delft University of Technology, The Netherlands \and \url{https://arnoudvanderleer.github.io/}}{arnoudvanderleer@gmail.com}{https://orcid.org/0009-0007-0458-3679}{}

\author{Kobe Wullaert}{Delft University of Technology, The Netherlands \and \url{https://kobewullaert.github.io/}}{K.F.Wullaert@tudelft.nl}{https://orcid.org/0000-0003-4281-2739}{}

\author{Benedikt Ahrens}{Delft University of Technology, The Netherlands \and \url{https://benediktahrens.gitlab.io/}}{B.P.Ahrens@tudelft.nl}{http://orcid.org/0000-0002-6786-4538}{}

\authorrunning{A.\,A. van der Leer, K. Wullaert and B. Ahrens}

\Copyright{Arnoud van der Leer, Kobe Wullaert, and Benedikt Ahrens}

\ccsdesc[500]{Theory of computation~Denotational semantics}
\ccsdesc[500]{Theory of computation~Logic and verification}

\keywords{Lambda calculi, algebraic theories, categorical semantics, Karoubi envelope, formalization, Rocq-UniMath, univalent foundations}
\category{}
\relatedversion{}

\acknowledgements{We thank Mohamed Barakat, Tom de Jong, and Niels van der Weide, as well as the anonymous referees for valuable feedback on versions of this paper.}

\usepackage{stmaryrd} %
\usepackage{tikz-cd}
\usepackage{csquotes} %
\usepackage[utf8]{inputenc}
\usepackage{listingsutf8}
\DeclareUnicodeCharacter{03BB}{\lambda}
\DeclareUnicodeCharacter{2022}{\bullet}
\newcounter{resume}

\newcommand{\iindex}[1]{\textbf{#1}}
\newcommand{\eemph}[1]{\emph{#1}}

\newcommand{\id}[1]{\ensuremath{\mathsf{id}_{#1}}}
\newcommand{\op}[1]{\ensuremath{#1^{\mathsf{op}}}}

\newcommand\cat[1]{\mathbf{#1}}
\newcommand\SET{\cat{Set}}
\newcommand\TYPE{\cat{Type}}
\newcommand\AlgTh{\cat{AlgTh}}
\newcommand\LamTh{\cat{LamTh}}
\newcommand\Pshf[1]{\cat{Pshf}_{#1}}

\newcommand\R{\cat{R}}
\newcommand\A{\cat{A}}

\newcommand\C{\cat{C}}
\newcommand\D{\cat{D}}

\renewcommand\L{\cat{L}}

\newcommand{\ReflOb}{\mathsf{RO}}
\newcommand{\LamThType}{\mathsf{LamTh}}
\newcommand\qand{\quad \text{and} \quad}

\newcommand{\trunc}[1]{\ensuremath{\| {#1} \|}}

\newcommand\setkaroubi[1]{\cat{R}^S(#1)}
\newcommand\univalentkaroubi[1]{\cat{R}^U(#1)}
\newcommand\monoidcat[1]{\C_{#1}}

\newcommand{\interp}[1]{\ensuremath{\llbracket {#1} \rrbracket }}

\DeclareFontFamily{U}{dmjhira}{}
\DeclareFontShape{U}{dmjhira}{m}{n}{ <-> dmjhira }{}

\DeclareRobustCommand{\yo}{\text{\usefont{U}{dmjhira}{m}{n}\symbol{"48}}}

\newcommand{\coqsymbol}{\begingroup\normalfont\includegraphics[height=\fontcharht\font`\{]{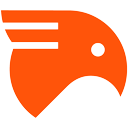}\endgroup}
\newcommand{\coqdocbaseurl}{https://arnoudvanderleer.github.io/UniMath-docs/5941f44/UniMath.}
\newcommand{\coqident}[2]{\href{\coqdocbaseurl #1.html\##2}{\coqsymbol}}
\newcommand{\coqidenturl}[3]{\href{\coqdocbaseurl #1.html\##3}{\coqsymbol}}
\newcommand{\coqfile}[1]{\href{\coqdocbaseurl #1.html}{\nolinkurl{#1}}}

\newtheorem*{theorem*}{Theorem}

\usepackage[draft]{optional}
\usepackage[colorinlistoftodos,prependcaption,textsize=tiny]{todonotes}%
\newcommand{\plan}[1]{}
\newcommand{\BA}[1]{}
\newcommand{\AL}[1]{}
\newcommand{\KW}[1]{}
\opt{draft}{
   \renewcommand{\plan}[1]{\textcolor{blue}{Plan: #1}\PackageWarning{TODO}{Plan}}
   \renewcommand{\BA}[1]{\todo[color=orange!30]{BA: #1}\PackageWarning{TODO}{BA: #1}}
   \renewcommand{\AL}[1]{\todo[color=green!30]{AL: #1}\PackageWarning{TODO}{AL: #1}}
   \renewcommand{\KW}[1]{\todo[color=red!30]{KW: #1}\PackageWarning{TODO}{KW: #1}}
}

\begin{document}

  \maketitle

  \begin{abstract}
    Lambek and Scott constructed a correspondence between simply-typed lambda calculi and Cartesian closed categories.
    Scott's Representation Theorem is a cousin to this result for \eemph{untyped} lambda calculi.
    It states that every untyped lambda calculus arises from a reflexive object in some category.

    We present a formalization of Scott's Representation Theorem in univalent foundations, in the (Rocq-)UniMath library.
    Specifically, we implement two proofs of that theorem, one by Scott and one by Hyland.
    We also explain the role of the Karoubi envelope --- a categorical construction --- in the proofs and the impact the chosen foundation has on this construction.
    Finally, we report on some automation we have implemented for the reduction of $\lambda$-terms.
  \end{abstract}

  \section{Introduction}

We present a formalization of two proofs of Dana Scott's Representation Theorem (SRT) and a detailed comparison between them, all incorporated into the UniMath library \cite{UniMath} of univalent mathematics, based on the proof assistant Rocq (formerly named Coq) \cite{Coq-refman}.

SRT provides denotational semantics for the untyped lambda calculus.
It is a cousin to the arguably better-known result by Lambek and P.J. Scott \cite{lambek-scott}, which provides denotational semantics for the \eemph{simply-typed} lambda calculus (\cref{subsec:lambek-correspondence}).
Specifically, SRT states that lambda calculi can be modelled by certain objects in certain categories, and that every lambda calculus arises as such an object (\cref{subsec:scott-intro}).

In this work, based on \cite{thesis-arnoud}, we formalize the original proof of SRT by Dana Scott \cite{scott} and a more modern proof by Martin Hyland \cite{Hyland}.
We formalize them in univalent foundations, which are more granular than traditional set-theoretic foundations. This shows strongly in our analysis of these proofs: univalent foundations provide exactly the right language to explain the fundamental difference between the two constructions.

In the remainder of the introduction,
we review Lambek and Scott's correspondence between Cartesian closed categories and simply-typed lambda calculi in \cref{subsec:lambek-correspondence}.
Afterwards, in \cref{subsec:scott-intro}, we give an introduction to SRT.
In \cref{subsec:univalence}, we provide a brief overview of univalent foundations and category theory therein, to the extent used in this paper.
In \cref{subsec:synopsis}, we give an overview of the paper and pointers to the formalization.

\subsection{The Correspondence Between Categories and Typed \texorpdfstring{$ \lambda $}{lambda}-calculi}\label{subsec:lambek-correspondence}

We sketch the correspondence between simply-typed lambda calculi (STLCs) and Cartesian closed categories.
A STLC is a simply-typed language with at least abstraction and application.
The type system of the language should be closed under function types $A \to B$.
Lambek and Scott \cite{lambek-scott} describe the construction of denotational models for such calculi.

\begin{definition}
  A \iindex{Cartesian closed category}, abbreviated \iindex{CCC}, is a category $\C$, equipped with a terminal object $1$, binary products $A \times B$, and exponentials $B^A$.
\end{definition}

Now, a Cartesian closed category $\C$ gives rise to a STLC, its ``internal language'': roughly, take the objects of the category to be types, and morphisms $f : A \to B$ of the category to be terms of type $B$ in context $A$.
Conversely, any simply-typed lambda calculus gives rise to a Cartesian closed category, as follows.
A type $A$ of the calculus gives an object $\interp{A}$ in the category.
A term $t : A$ of type $A$ in context $x_1:A_1,\ldots,x_n:A_n$ gives a morphism $\interp{t} : \interp{A_1} \times\ldots \times \interp{A_n} \to \interp{A}$.

Internal logic and interpretation, together, provide a \eemph{correspondence between simply-typed lambda calculi and Cartesian closed categories}:
\begin{equation}\label{diag:stlc}
\begin{tikzcd}[column sep = large]
{\mathsf{STLC}} \arrow[r, bend left, "\text{interpretation}"]
& {\mathsf{CCC}} \arrow[l, bend left, "\text{internal logic}"]
\end{tikzcd}
\end{equation}

\subsection{Scott's Representation Theorem}
\label{subsec:scott-intro}

In 1969, Scott gave a famous series of lectures \cite{SCOTT1993411} where he initially claimed that untyped $\lambda$-calculi did not have (non-trivial) categorical models.
Intuitively, for an object $D$ of some category to be a model of the untyped lambda calculus, one would need functions $D \leftrightarrows D^D$ modelling abstraction and application, such that the composition $D^D \to D \to D^D$ is the identity --- to model $\beta$-equality.
(For $\eta$-equality, the other composite must be the identity on $D$.)
In particular, this would mean that the function space $D^D$ can be embedded into $D$ itself.
In the category of sets, this is only possible for $D = 1$.
However, as Scott's lectures progressed, he realized that his initial claim was incorrect.
In the end, Scott constructed a model $\mathcal{D}_\infty$ in the category of directed complete partial orders (dcpos).
Later, models such as \eemph{Scott's graph model} and \eemph{Böhm's tree model} were constructed; see, e.g., \cite[Part~V]{DBLP:books/daglib/0067558}.

To arrive at a result similar to the correspondence of Diagram \eqref{diag:stlc}, SRT abstracts away from particular calculi and models.
We use Hyland's \cite{Hyland} axiomatization of lambda calculi as $\lambda$-theories, which are particular \eemph{algebraic theories} from universal algebra.
A $\lambda$-theory is an algebraic theory equipped with operations modelling abstraction and application, satisfying $\beta$- (and potentially $\eta$-) equality.
On the semantic side, a model of a lambda calculus is defined to be a \eemph{reflexive object} $D$ in a Cartesian closed category; that is, a diagram $D \leftrightarrows D^D$ that exhibits $D^D$ as a retract of $D$ (see \cref{dfn:reflexive-obj}).

For every reflexive object, one can construct its so-called \emph{endomorphism theory}.
Then SRT can be stated as follows: every untyped $\lambda$-calculus arises as the  endomorphism theory of a reflexive object.
To summarize, SRT provides the following diagram, where $\LamThType$ consists of $\lambda$-theories and $\ReflOb$ of pairs of a Cartesian closed category and a reflexive object therein:
\begin{equation}\label{diag:intro-diag}
\begin{tikzcd}[column sep = large]
{\LamThType} \arrow[r, bend left, "\mathrm{interpretation}"] & {\ReflOb} \arrow[l, bend left, swap, "\mathrm{endom.~theory}"']
\end{tikzcd}
\end{equation}
Unlike in the case of \eemph{simply-typed} lambda calculi, this diagram does not represent an equivalence, but only a \eemph{section-retraction pair}.
That is, if we start with a lambda calculus, interpret it and then take the endomorphism theory of the interpretation, we end up with ``the same'' lambda calculus.
However, starting with a reflexive object, taking its endomorphism theory and then the interpretation does not necessarily yield the original reflexive object; see also \cref{rem:not-inj}.

Scott and Hyland gave different constructions of the interpretation of a lambda calculus and the proof that this forms a section.
In this paper, we formalize both, and compare them, in the light of univalent foundations.

\subsection{A Quick Tour of Univalent Foundations and Univalent Categories}
\label{subsec:univalence}

In this section, we give a brief introduction to univalent foundations and univalent category theory.
Other short introductions can be found in \cite{grayson-intro,AhrensNorth}.
A comprehensive introduction is given in the HoTT book \cite{hottbook}; in particular, \cite[Chap.~9]{hottbook} discusses univalent categories.

Univalent foundations (UF) are foundations of mathematics satisfying the univalence principle.
This principle says, roughly, that two equivalent mathematical objects satisfy all the same properties.
Technically, UF are an extension of Martin-Löf type theory (MLTT) \cite{martin-lof-type-theory} by the univalence axiom and, optionally, (certain) higher inductive types.
We assume the reader is familiar with the basics of MLTT as discussed, for instance, in \cite[Chap.~1]{hottbook}.
To fix notation: we write $ g \circ f $ for the composition of functions $ f: X \to Y $ and $ g: Y \to Z $.

Given a type $ X $ and elements $ x, y: X $, we denote by $x = y$ the identity type between $x$ and $y$.
The type $x = y$ can contain multiple distinct elements, that is, different identity terms from $ x $ to $ y $.
For any $x : X$, we have the identity term $\mathsf{refl}(x) : x = x$.
In particular, given two types $S$ and $T$ (elements of some universe type), we have the type $S = T$.
There is also a seemingly weaker notion of sameness, called \eemph{equivalence}, written $S \simeq T$;
its elements are functions $S \to T$ with a two-sided inverse (see \cite[Chap.~4]{hottbook} for details).
We have a function $\mathrm{idtoequiv}: (S = T) \to (S \simeq T)$,
sending $\mathsf{refl}(S)$ to the identity equivalence on $S$.
The \iindex{univalence axiom} says that this map is an equivalence, for any $S$ and $T$.
In particular, every structure on types transports along an equivalence of types.

In univalent foundations, types are \eemph{stratified} according to how ``complex'' their identity types are: Let $ T $ be a type.
It is \iindex{contractible} if there is an equivalence $e : T \simeq 1$ to the unit type $1$.
It is a \iindex{proposition} if any two of its elements are identical: $\prod_{s,t : T} s = t$.
It is a \iindex{set} if all of its identity types $s = t$, for $s,t : T$, are propositions (identity proofs in $ T $ are unique).
It is a \iindex{groupoid} if all of its identity types $s = t$, for $s,t : T$, are sets.
This stratification of types can be continued recursively, but the aforementioned definitions suffice for the purpose of this paper.

Given a type $A$, we form a new type $\trunc{A}$, the \iindex{propositional truncation of $A$};
$\trunc{A}$ is a proposition, and inhabited if and only if $A$ is inhabited.
When postulating that something ``merely exists in $T$ with property $P$'' (such as in \cref{def:karoubi''}), we formalize it as $\trunc{\sum_{x:T} P(x)}$.

There are two notions of \iindex{category} (\coqident{CategoryTheory.Core.Categories}{category}) in univalent foundations.
Both kinds consist, in particular, of a type $O$ of objects and a dependent type $M : O \to O \to \TYPE$ of morphisms.
Given a category $\C$, we write $ \C(X, Y) $ or $ X \to Y $ for the type of morphisms between objects $ X $ and $ Y $.
A category also has suitably typed composition and identity operations, and proofs of the associativity and unitality of composition.
To ensure that unitality and associativity are propositions, we require that all the $\C(X,Y)$ are sets.
Just like with functions, we write $ g \circ f $ for the composition of morphisms $ f: \C(X, Y) $ and $ g: \C(Y, Z) $.
For details, we refer to \cite[Def.~3.1]{univalent-categories} --- there, categories are called \iindex{pre}categories.

A (pre)category as described above does not correspond to a category in Voevodsky's model of univalent foundations in simplicial sets \cite{kapulkin2021simplicial}; it has data that is not usually present in a category, namely a potentially non-trivial identity type on the type $O$ of objects.
There are two ways to ``kill'' that data.
Firstly, we could assert that the type $O$ of objects should also form a set in the above sense; this leads to the notion of \iindex{setcategory}.
Secondly, we could assert that the identities $a = b$ of $a,b:O$ correspond exactly to \iindex{isomorphisms} $a \cong b$ of the category; this leads to the notion of \iindex{univalent category}.
More precisely, in a univalent category $\C$ we demand that for every $a,b : \C$, the map
$\mathsf{idtoiso}_{a,b} : (a = b) \to (a \cong b)$,
mapping $\textsf{refl}(a)$ to the identity isomorphism on $a$, is an equivalence, in analogy to the univalence axiom.
One can show that the type of objects of a univalent category forms a groupoid.

There are thus two theories of categories in univalent foundations.
Setcategory theory behaves much like traditional category theory; however, many naturally occurring categories are not setcategories.
On the other hand, there are many univalent categories.
Importantly, the category $\SET$ of sets (of a fixed universe) in the sense described above, is univalent; its objects are pairs of a type $X$ (of a fixed universe) together with a proof that $X$ is a set, and morphisms are type-theoretic functions.
Roughly, univalence of this category is a consequence of the univalence axiom (for that universe).
Furthermore, for two categories $\C$ and $\D$, the functor category $[\C,\D]$ is univalent if $\D$ is univalent; hence, in particular, presheaf categories $P \A = [\op \A, \SET]$ are univalent.
Still, not every category is univalent, e.g., the category $\bullet \leftrightarrows \bullet$ with two distinct but isomorphic objects, or the set-Karoubi envelope of \cref{subsec:set-karoubi}, see \cref{ex:karoubi-not-univalent}.
However, every category has a univalent replacement, which we call its \iindex{Rezk completion}:

\begin{theorem}[{\cite[Thm.~8.5]{univalent-categories}}]
  For every category $ \C $, there exists a univalent category $ \D $ with a fully faithful and essentially surjective functor $\iota: \C \hookrightarrow \D$.
\end{theorem}

\textbf{In summary}, univalent foundations give us two flavors of category theory: category theory up to isomorphism (setcategories) and category theory up to adjoint equivalence (univalent categories).
We demonstrate the difference between Scott's and Hyland's proof of SRT by showing that Scott's construction happens in the realm of setcategories, whereas Hyland's construction happens in the realm of univalent categories. Furthermore, Hyland's construction yields the Rezk completion of Scott's construction.

\subsection{Synopsis and Guide to the Formalization}
\label{subsec:synopsis}
In \cref{sec:SRT} we present our formalization of SRT as shown in \cref{diag:intro-diag}. More precisely, we formalize two different constructions of the function labelled ``interpretation'', one by Scott, and one by Hyland,
and we prove that both are sections to the function associating to a reflexive object its endomorphism theory.
In \cref{sec:karoubi} we zoom in on the different interpretation functions.
Both use a variant of the \eemph{Karoubi envelope} of a category.
We provide a novel analysis and comparison of the two variants, through the lens of univalent foundations, and show how the two variants relate Scott's and Hylands constructions.
In \cref{sec:tactic} we describe a tactic for the manipulation of lambda terms.
In \cref{sec:related-work} we discuss related work.

Most of the definitions and results presented in this paper are formalized and computer-checked in UniMath \cite{UniMath}, a library of univalent mathematics based on Rocq.
Our code has been integrated into the UniMath library, and comprises more than 11000 lines of code, split up more or less evenly between definitions and proofs.
We can identify the following main components: the definitions of the basic categories with their object and morphism types (3000 lines); the proof of Scott's version of SRT, which uses a lot of reasoning about $ \lambda $-terms (3000 lines); the definition of the Karoubi envelope (2000 lines); the constructions of various algebraic theories, like the endomorphism theory, and the proof of the relation between the representation theorems (both 1000 lines).

Throughout the paper, definitions and results are decorated with links, for example, \coqident{AlgebraicTheories.Examples.FreeTheory}{free_theory}, to an HTML version of UniMath.
That HTML version is derived from commit \href{https://github.com/UniMath/UniMath/tree/5941f44e3c13d2ac385f5a8b9b486c0088dd3f46}{5941f44} of UniMath.
Proof-checking and creation of the HTML documentation can be reproduced locally by following
the UniMath compilation instructions.
Note that for technical reasons, the formalization does not always closely follow the material in this paper, with the main difference described in \cref{rem:displayed-cats-obfuscation}.
Another point of complication is the necessary switching between two different ways to encode n-tuples $ x: X^n $, either as nested 2-tuples $ ((\dots ((x_1, x_2), x_3), ...), x_n) $ or as functions $ x: \{1, \dots, n\} \to X $.

  \section{What we Formalized: Scott's Representation Theorem}
\label{sec:SRT}

In this section, we present our formalization of SRT.
In \cref{subsec:algebraic-theories}, we define the type $\LamThType$ of $\lambda$-theories, the type $\ReflOb$ of reflexive objects,
and the endomorphism theory generated by a reflexive object, in the form of a function $E: \ReflOb \to \LamThType$
($E$ for ``endomorphism theory'').
In \cref{subsec:scott,subsec:Hyland}, we present Scott's and Hyland's constructions of a section to $E: \ReflOb \to \LamThType$, that is, of functions $S,H : \LamThType \to \ReflOb$ (where $S$ and $H$ stand for ``Scott'' and ``Hyland'', respectively) such that $E \circ S = \id{\LamThType} = E \circ H$.
\begin{equation}%
\begin{tikzcd}[column sep = large]
{\LamThType} \arrow[r, bend left = 30, "S"'] \ar[r,bend left = 45, "H"] & {\ReflOb} \arrow[l, bend left, swap, "E"']
\end{tikzcd}
\end{equation}
\begin{remark}\label{rem:section-no-iso}
  It might seem surprising that we can show $E \circ S = \id{\LamThType}$ instead of just $E (S (T)) \cong T$ for $T : \LamThType$ and a suitable notion of isomorphism of $\lambda$-theories.
  We can show the seemingly stronger identity thanks to the univalence axiom.
  Specifically, the univalence axiom entails that isomorphisms $T \cong T'$ of $\lambda$-theories coincide with identities $T = T'$ --- expressed below by the statement that the category of $\lambda$-theories is univalent (\cref{prop:lamth-univalence}).
  That is, our statement of SRT does not make reference to (iso)morphisms of $\lambda$-theories; the construction, however, does: we construct an isomorphism first, and turn it into an identity using univalence afterwards.
\end{remark}

\subsection{ Untyped \texorpdfstring{$ \lambda $}{Lambda}-Calculi and Their Denotational Semantics}
\label{subsec:algebraic-theories}

We give a definition of untyped \eemph{$\lambda$-calculi} (\cref{def:lambda-theory}), an extension of the concept of an \eemph{algebraic theory} from universal algebra (\cref{def:algebraic-theory}).
We also define reflexive objects in Cartesian closed categories, the intended denotational semantics for such calculi.

\begin{definition}[\coqident{AlgebraicTheories.AlgebraicTheories}{algebraic_theory}]\label{def:algebraic-theory}
  We define an \iindex{algebraic theory} $ T $ to be a sequence of sets $ T_n $ indexed over $ \mathbb N $ with for all $ 1 \leq i \leq n $ elements (``variables'' or ``projections'') $ x_{n, i}: T_n $ (we often leave $ n $ implicit and write $ x_i $), together with a substitution operation $\_ \bullet \_: T_m \times T_n^m \to T_n$
  for all $ m $ and $ n $, such that for all $ 1 \leq j \leq l $, $ f: T_l $, $ g: T_m^l $ and $ h: T_n^m $,
  \[
    x_j \bullet g = g_j, \quad
    f \bullet (x_{l, i})_i = f \qand
    (f \bullet g) \bullet h = f \bullet (g_i \bullet h)_i.
  \]
\end{definition}

Algebraic theories are also known as \emph{(abstract) clones}; they are similar to \emph{operads} --- see, e.g., \cite{lambda-monoid, tronin} for details.

\begin{definition}[\coqident{AlgebraicTheories.AlgebraicTheoryMorphisms}{algebraic_theory_morphism}]
  A \iindex{morphism $ f $ between algebraic theories} $ T $ and $ T^\prime $ is a sequence of functions $ f_n: T_n \to T^\prime_n $ (we usually leave the $ n $ implicit and just write $ f $ if the context is clear) such that for all $ 1 \leq j \leq n $, $ s: T_m $ and $ t: T_n^m $, $f_n(x_j) = x_j$, and $f_n(s \bullet t) = f_m(s) \bullet (f_n(t_i))_i$.
\end{definition}

Algebraic theories and their morphisms form a category \iindex{$ \AlgTh $} (\coqident{AlgebraicTheories.AlgebraicTheoryCategoryCore}{algebraic_theory_cat}).
Furthermore, an isomorphism $S \cong T$ of algebraic theories consists of pointwise bijections $f_n : S_n \cong T_n$ that respect the variables and substitution.
By univalence, each $f_n$ corresponds to an identity $p_n : S_n = T_n$ also respecting the algebraic structure.
Again by univalence, the $p_n$'s combine into an identity of algebraic theories $S = T$.
Hence:

\begin{proposition}[\coqident{AlgebraicTheories.AlgebraicTheoryCategory}{is_univalent_algebraic_theory_cat}]
\label{prop:algth-univ}
  $ \AlgTh $ is univalent.
\end{proposition}

\begin{example}[\coqident{AlgebraicTheories.Examples.FreeTheory}{free_theory}\coqident{AlgebraicTheories.Examples.FreeTheory}{free_functor_is_free}]
We have a \eemph{forgetful} functor from $\AlgTh$ to $\SET$ sending an algebraic theory $T$ to the set $T_0$.
Conversely, for every set $A$ one can construct an algebraic theory $T(A)$ with $T(A)_n =  \{1,\ldots, n\} \sqcup A$.
The variables are given by $x_{n,i} = i : T(A)_n$, for $0 < i \leq n$.
The substitution $f \bullet g$ is defined as follows:
if $f : A$, then $f \bullet g = f$;
if $1 \leq f \leq n$, then $f \bullet g = g_f$.
Then, $A \mapsto T(A)$ is part of a functor $T : \SET \to \AlgTh$.
Furthermore, $T$ is a left adjoint to the forgetful functor.
\end{example}

Let $ \iota_{m, n} : T_m \to T_{m + n} $ be the ``weakening'' function $ f \mapsto f \bullet (x_{m + n, 1}, \dots, x_{m + n, m}) $. Note that $\iota_{m, n}(f) \bullet g = f \bullet (g_i)_{i \leq m}$ and $\iota_{m, n}(f \bullet g) = f \bullet (\iota_{m, n}(g_i))_i$.

\begin{definition}[\coqidenturl{AlgebraicTheories.LambdaTheories}{beta_lambda_theory}{7819fa87f50e6d0a798b45a0a93dc7ce}]\label{def:lambda-theory}
  A \iindex{$ \lambda $-theory} is an algebraic theory $ L $, together with sequences of functions $ \lambda_n: L_{n + 1} \to L_n $ and $ \rho_n: L_n \to L_{n + 1} $, such that for all $ f: L_{m + 1} $, $ g: L_m $ and $ h: L_n^m $,
  \[
    \lambda_m(f) \bullet h = \lambda_n(f \bullet ((\iota_{n, 1}(h_i))_i + (x_{n + 1}))) \qand
    \rho_n(g \bullet h) = \rho_m(g) \bullet ((\iota_{n, 1}(h_i))_i + (x_{n + 1})).
  \]
  Furthermore, we assume that $L$ satisfies $\beta$-equality, that is $\rho_n \circ \lambda_n = \id{L_n} $ for all $ n $.
\end{definition}

We also formalized the notion of $ \lambda $-theory without $ \beta $-equality, and of $ \lambda $-theory with both $ \beta $- and $ \eta $-equality (meaning $\lambda_n \circ \rho_n = \id{L_n} $), but we only formalized the main results for the $\lambda$-theories of \cref{def:lambda-theory}.

\begin{example}[\coqidenturl{AlgebraicTheories.Examples.LambdaCalculus}{lambda_calculus_has_beta}{1d6f6b93b532caf94fefa421e6c75e82}]\label{ex:lambda-calculus-without-constants}
  An important example of a $ \lambda $-theory is provided by the $ \lambda $-calculus without constants $ \Lambda $. We present it here as a higher inductive type, to have it satisfy $ \beta $-equality. For $ i : \{ 1, \dots, n\} $, $ s, t : \Lambda_n $, $ u : \Lambda_{n + 1} $ and $ v_1, \dots, v_n : \Lambda_m $, it has constructors
  \[
    \mathtt{Var}_n(i) : \Lambda_n, \quad
    \mathtt{App}_n(s, t) : \Lambda_n, \quad
    \mathtt{Abs}_n(u) : \Lambda_n \qand
    \mathtt{Subst}_{n, m}(s, (v_1, \dots, v_n)) : \Lambda_m,
  \]
  together with identities about the interaction between $ \mathtt{Subst} $ with every constructor, as well as the identity $ \mathtt{App}(\mathtt{Abs}(f), g) = \mathtt{Subst}(f, (\mathtt{Var}(1), \dots, \mathtt{Var}(n), g)) $ for $ \beta$-equality.
   The functions $ \rho_n : \Lambda_n \to \Lambda_{n + 1} $ are given by $ \rho(f) = \mathtt{App}(\mathtt{Subst}(f, (x_1, \dots, x_n)), x_{n + 1}) $.
   \lipicsEnd
\end{example}

\begin{remark}[\coqident{AlgebraicTheories.LambdaCalculus}{lambda_calculus}]
  Because the UniMath community does not permit the usage of inductive types in the core library, we defined $ \Lambda $ \emph{synthetically}, instead of constructing it: our formalization takes a hypothesis $ \Lambda $ of a type describing the $ \lambda $-calculus without constants as a sequence of sets $ (\Lambda_n)_n $ with functions $ \mathtt{Var} $, $ \mathtt{App} $, $ \mathtt{Abs} $ and $ \mathtt{Subst} $ that satisfy the correct identities. The type of $ \Lambda $ also contains an \emph{induction principle}, which allows one to construct terms of type $ \prod_n \prod_{f : \Lambda_n} T_n(f) $ for a family of sets $ T $.
  The non-dependent version of this induction principle takes a set $ X $ and functions $ var_n: \{1, ..., n\} \to X $, $ app_n: X \times X \to X $, $ abs_n : X \to X $ and $ subst_n: X \times X^n \to X $, satisfying the same identities as the constructors of $ \Lambda $, and gives functions $ f_n : \Lambda_n \to X $ such that $ f(\mathtt{Var}(i)) = var(i) $, $ f(\mathtt{App}(s, t)) = app(f(s), f(t)) $ etc.
\end{remark}

\begin{definition}[\coqidenturl{AlgebraicTheories.LambdaTheoryMorphisms}{beta_lambda_theory_morphism}{f0f77d9681a5ba6c674ab8f76631407c}]
  A \iindex{morphism $ f $ between $ \lambda $-theories} $ L $ and $ L^\prime $ is an algebraic theory morphism $ f $ such that $f_n \circ \lambda_n = \lambda_n \circ f_{n+1}$ and $\rho_n \circ f_n = f_{n+1} \circ \rho_n$.
\end{definition}

$ \lambda $-theories and their morphisms form a category \iindex{$ \LamTh $} (\coqidenturl{AlgebraicTheories.LambdaTheoryCategoryCore}{beta_lambda_theory_cat}{0a7d3a86217f13daaed748d960f50346})
with $\LamThType$ as the underlying type of objects.

\begin{example}[\coqident{AlgebraicTheories.Examples.LambdaCalculus}{lambda_calculus_is_initial}]
  For any $ \lambda $-theory $ L $, there is a unique morphism $ i : \Lambda \to L $, defined using structural induction on the terms of $ \Lambda $: it sends variables to variables, substitution to substitution, etc. This makes $ \Lambda $ into the initial object of $ \LamTh $.
\end{example}

\noindent
We can prove the following analogously to \cref{prop:algth-univ}:
\begin{proposition}[\coqidenturl{AlgebraicTheories.LambdaTheoryCategory}{is_univalent_beta_lambda_theory_cat}{8f422c1ffa4cf80ba239c99cb728d551}]
\label{prop:lamth-univalence}
  $ \LamTh $ is univalent.
\end{proposition}

\Cref{prop:lamth-univalence} expresses that identities $T = T'$ of $\lambda$-theories are equivalent to isomorphisms $T \cong T'$;
this enables us to express SRT purely on the level of types, without needing to mention (iso)morphisms of $\lambda$-theories, as discussed in \cref{rem:section-no-iso}.

Let $ L $ be a $ \lambda $-theory. Hyland shows that $ \rho_n(f) $ corresponds to $ f x_{n+1} $, which is $ f $ applied to $ x_{n+1} $ \cite[Section 3.1]{Hyland}. Now consider the element $ \rho(x_{1, 1}) : L_2 $.
\begin{definition}[\coqident{AlgebraicTheories.LambdaTheories}{app'}]
  Using the substitution, we have binary operations on the $ L_n $, sending $ (f, g) : L_n \times L_n $ to $ \rho(x_{1, 1}) \bullet (f, g) : L_n $.
\end{definition}
\begin{remark}\label{lambda-calculus-operations}
  This means that $ L $ has the structure of a $ \lambda $-calculus with $ \beta $-equality, with variables $ x_i $, application $ f g = \rho(x_{1, 1}) \bullet (f, g) $ and abstraction $ \lambda x_{n+1}, f = \lambda(f) $.
\end{remark}

Now we are ready to talk about denotational semantics for the untyped $ \lambda $-calculus, for which we will need the following definition.
\begin{definition}[\coqident{AlgebraicTheories.ReflexiveObjects}{reflexive_object}]
\label{dfn:reflexive-obj}
  If $ \C $ is a category with binary products, a \iindex{reflexive object} in $ \C $ is an object $ X $ such that the exponential $ X^X $ exists, and with a retraction from $ X $ onto $ X^X $: morphisms $ f: X \to X^X $ and $ g : X^X \to X $ such that $f \circ g = \id{X^X} $.
\end{definition}

\begin{definition}[\coqident{AlgebraicTheories.Examples.EndomorphismTheory}{reflexive_object_to_lambda_theory}]\label{def:endomorphism-theory}
  We define a function $E : \ReflOb \to \LamThType$ as follows.
  Let $ X $ be a reflexive object in a category $ \C $, with the retraction given by $ \mathrm{abs}: X^X \to X $ and $ \mathrm{app}: X \to X^X $.

  The \iindex{endomorphism theory} $ E(X) $ of $ X $ is the algebraic theory given by $ E(X)_n = \C(X^n, X) $ with projections as variables $ x_{n, i}: X^n \to X $ and a substitution that sends $ f: X^m \to X $ and $ g_1, \dots, g_m: X^n \to X $ to $f \circ \langle g_i \rangle_i: X^n \to X $.

  If $ \varphi_Y : \C(X \times Y, X) \xrightarrow{\sim} \C(Y, X^X) $ is the bijection given by the exponential $ X^X $, we can give $ E(X) $ a $ \lambda $-theory structure by setting, for $ f: E(X)_{n + 1} $ and $ g: E(X)_n $,
  \[
    \lambda(f) = \mathrm{abs} \circ \varphi_{X^n}(f) \qand
    \rho(g) = \varphi_{X^n}^{-1}(\mathrm{app} \circ g).
  \]

  $ \beta $-equality for $ E(X) $ follows immediately from the fact that $ \mathrm{app} : X \to X^X $ is a retraction.
\end{definition}

The proofs that $ E(X) $ is an algebraic theory follow mainly from properties of the product and naturality of $ \varphi_Y $. Note that $ E(X) $ would satisfy $ \eta $-equality if $ \mathrm{app} $ were a section.

\begin{example}
\label{exa:endo-th-of-unit}
  As a trivial example, take $ X = \{ \star \} $ in the category of sets. The set of functions $ X^X = X \to X $ is isomorphic to $ X $, and so we get an endomorphism theory $ E(X) $ with $ E(X)_n = \{ \star \} $.
\end{example}

Because of cardinality reasons, \cref{exa:endo-th-of-unit} is the only reflexive object in the category of sets. Hence, for nontrivial models of the untyped $\lambda$-calculi, we have to look in categories different from $\SET$.
Scott's model $D_\infty$ lives in the category of dcpos whose morphisms are (continuous) section-retraction pairs \cite{scott-continuous}.

SRT states that we can represent every untyped $ \lambda $-calculus as an endomorphism theory of some reflexive object:
\begin{theorem*}
  The function $ E $ of \cref{def:endomorphism-theory} has a right inverse:
  \begin{equation}\label{diag:srt}
    \begin{tikzcd}[column sep = large]
      {\LamThType} \arrow[r, bend left, "\mathrm{interpretation}"] & {\ReflOb} \arrow[l, bend left, swap, "E"']
    \end{tikzcd}
  \end{equation}
\end{theorem*}

In the following two subsections, we will discuss two proofs of this theorem: the original proof by Dana Scott (\cref{Scott}) and a more abstract proof by Martin Hyland (\cref{thm:Hyland}).

\begin{remark}[\coqident{AlgebraicTheories.RepresentationTheoremsRelation}{endomorphism_theory_no_left_inverse}]\label{rem:not-inj}
  The constructions by Scott and Hyland yield different reflexive objects. In particular, Hyland's category has an empty object with no morphisms into it, whereas Scott's category always has at least one morphism between any two objects. By the representation theorems, the function $ E $ sends both of these to the same $ \lambda $-theory, so $ E $ is not injective. Therefore, contrary to the situation for STLCs in \cref{subsec:lambek-correspondence}, $ E $ is not an equivalence.
\end{remark}

\begin{remark}
  By definition, the $\lambda$-theories we consider satisfy $\beta$-equality.
  A variant of SRT also holds for $\lambda$-theories satisfying both $\beta$ and $\eta$-equality.
\end{remark}

\begin{remark}\label{rem:displayed-cats-obfuscation}
  We use the technology of \emph{displayed categories} \cite{displayed-categories} for implementing many of the concepts in this section.
  A displayed category $ \D $ over a category $ \C $ gives the objects and morphisms of $ \C $ ``additional structure''. Consider how a $ \lambda $-theory is an algebraic theory, together with operations $ \rho $ and $ \lambda $, and a $ \lambda $-theory morphism is an algebraic theory morphism, that preserves $ \rho $ and $ \lambda $. This is formalized by defining $ \LamTh $ to be displayed over $ \AlgTh $. In turn, $\AlgTh$ is displayed over $\SET^{\mathbb{N}}$.
  The goal of displayed categories is to build categories of complicated objects and morphisms in ``layers'', incrementally adding more structure. These layers can be built and reasoned about modularly.
  The proof of univalence of those categories, \cref{prop:algth-univ,prop:lamth-univalence}, is then especially simple: it relies on univalence of $\SET$ and univalence of the layers leading up to $\LamTh$.
  In the formalization, after the categories have been defined, we define the object types in terms of their categories, to minimize the amount of duplicate code. For example, we define a $ \lambda $-theory to be an object of $ \LamTh $.
  A drawback to this approach is that it somewhat obfuscates the precise definition of the objects and morphisms of such a category: the structure of a $ \lambda $-theory is hidden away in a stack of four displayed categories, although the constructor and accessors give some hints about what a $ \lambda $-theory consists of.
\end{remark}

\subsection{Scott's Construction}
\label{subsec:scott}
In this subsection, we will give Scott's original proof \cite{scott} of SRT.

It is a fairly syntactical proof, making heavy use of the operations of the $ \lambda $-calculus for the various constructions. We first define the category $ \R $ (\cref{R}) and show that it has products and exponential objects. Then we construct an object in this category and show that it is a reflexive object. Lastly, we construct the endomorphism theory and show that it is isomorphic to the $ \lambda $-theory that we started with (\cref{Scott}).

Let $ L $ be a $ \lambda $-theory. First of all, for $ a_1, a_2: L_0 $, we define
\begin{align*}
  a_1 \circ a_2 &= \lambda x_1, a_1 (a_2 x_1); &
  \pi_i &= \lambda x_1, x_1 (\lambda x_2 x_3, x_{i + 1});\\
  (a_1, a_2) &= \lambda x_1, x_1 a_1 a_2; &
  \langle a_1, a_2 \rangle &= \lambda x_1, (a_1 x_1, a_2 x_1).
\end{align*}
Although, actually, since every one of these starts with a $ \lambda $-abstraction, we need to lift the constants $ a_i $ to $ \iota_{0, 1}(a_i): L_1 $ to make the definitions above typecheck.
Note that $ \pi_i (a_1, a_2) = a_i $ and $ \pi_i \circ \langle a_1, a_2 \rangle = \lambda x_1, a_i x_1 $.

We define
$(a_i)_i = (a_1, \dots, a_n) = ((\dots(((\lambda x_1, x_1), a_1), a_2), \dots), a_n)$
(analogous for $\langle a_i \rangle_i$)
and correspondingly $\pi_{n, i} = \pi_2 \circ {\pi_1}^{n - i}$.

Scott introduces the following category, referred to as the \iindex{category of retracts}:
\begin{definition}[\coqident{AlgebraicTheories.CategoryOfRetracts}{R}]\label{R}
  Let $ \R $ be the category with as objects the $ A : L_0 $ such that $ A \circ A = A $ and as morphisms $ f : A \to B $ the $ f : L_0 $ such that $ B \circ f = f = f \circ A $. The composition is given by the composition of $ \lambda $-terms and the identity on $ A : \R $ is given by $ A $ itself.
\end{definition}

$ \R $ has a terminal object, given by $ I = \lambda x_1 x_2, x_2 : \R $ (\coqident{AlgebraicTheories.CategoryOfRetracts}{R_terminal}). It is terminal because for $ f: A \to I $, we have $ f = I \circ f = I $.

$ \R $ also has binary products $A_1 \times A_2 = \langle p_1, p_2 \rangle$ for $p_i = A_i \circ \pi_i$ the projections, with product morphisms $ \langle f, g \rangle $ (\coqident{AlgebraicTheories.CategoryOfRetracts}{R_binproducts}).

For $ B, C : \R $, we have an exponential object $ C^B = \lambda x_1, C \circ x_1 \circ B $, with a bijection $ \psi: \R(A \times B, C) \xrightarrow \sim \R(A, C^B) $, given by $ \psi(f) = \lambda x_1 x_2, f (x_1, x_2) $ and $ \psi^{-1}(f) = \lambda x_1, g (\pi_1 x_1) (\pi_2 x_1) $ (\coqident{AlgebraicTheories.CategoryOfRetracts}{R_exponentials}).

Now we can define our reflexive object:
\begin{definition}
[\coqident{AlgebraicTheories.CategoryOfRetracts}{U}\coqident{AlgebraicTheories.CategoryOfRetracts}{R_retraction_is_retraction}]
\label{def:Scott-reflexive}
  We define $ U : \R $, given by the identity $ \lambda x_1, x_1 $. For all $ A: \R $, we can also consider $ A $ as both a morphism $ r_A : \R(U, A) $ and a morphism $ s_A : \R(A, U) $ with $ r_A \circ s_A = \id A $, exhibiting $ A $ as a retract of $ U $. In particular, $U$ is reflexive.
\end{definition}

\noindent
This allows us to give Scott's proof of the representation theorem:

\begin{theorem}[\coqident{AlgebraicTheories.OriginalRepresentationTheorem}{endomorphism_theory_right_inverse}]\label{Scott}
  The function $ E $ of \cref{def:endomorphism-theory} has a right inverse $ S $, given by $ S(L) = (\R, U) $.
\end{theorem}
\begin{proof}
  Let $ L $ be a $ \lambda $-theory. As mentioned in \cref{def:Scott-reflexive}, $ U $ is a reflexive object, so the endomorphism theory $ E(U) $ has a $ \lambda $-theory structure.

  We have bijections $ \psi_n : E(U)_n \xrightarrow \sim L_n $, given by $ \psi_n(f) = \iota_{0, n}(f)(x_i)_i, $ and $ \psi^{-1}_n(g) = \lambda x_1, g \bullet (\pi_{n, i} x_1)_i $. These are bijections because $ E(U)_n = \{ f: L_0 \mid f \circ U^n = f \} $, with $ U^n = \langle \pi_{n, i} \rangle_i $.
  Explicitly, $ E(U) $ has variables $ x_{n, i} = \pi_{n, i} $, substitution $ f \bullet g = f \circ \langle g_i \rangle_i $, abstraction $ \lambda(f) = \lambda x_1 x_2, \iota_{0, 2}(f)(x_1, x_2) $ and application $ \rho(g) = \lambda x_1, \iota_{0, 1}(g) (\pi_1 x_1) (\pi_2 x_1) $ for $ f: E(U)_{n+1} $ and $ g: E(U)_n $. Using this, it is pretty straightforward to check that the $ \psi_n $ respect the $ \lambda $-theory structure, so $ \psi $ is an isomorphism of $ \lambda $-theories.
Then, univalence of $\LamTh$ gives an identity $ E(S(L)) = L $.
\end{proof}

\subsection{Hyland's Construction}
\label{subsec:Hyland}
While Scott proves the representation theorem in a very syntactical way, constructing the category $ \R $ for his representation theorem using the $ \lambda $-calculus operations, Hyland uses more machinery from category theory, and works with a different  category: the category of ($T$-)presheaves $\Pshf{T}$ of a $ \lambda $-theory $T$.

\begin{remark}
The $T$-presheaves can be described as the presheaves over the category $\L$, referred to as the \eemph{Lawvere theory} associated to $T$ (\cref{lem:lawvere-clone}).
Instead of working with the general construction, we spell out what this means, and refer to presheaves as the unfolded version hereof.
In \cref{Lawvere-presheaf}, we show that these notions coincide.
\end{remark}

In this section, we define the category of presheaves (\cref{presheaf}), and construct a reflexive object herein \cref{cor:Hyland-reflexive}.
Furthermore, we show that the endomorphism theory of the reflexive object is isomorphic to the $ \lambda $-theory that we started with (\cref{thm:Hyland}).

\begin{definition}[\coqident{AlgebraicTheories.Presheaves}{presheaf}]\label{presheaf}
  A \iindex{presheaf} $ P $ for an algebraic theory $ T $ is a sequence of sets $ P_n $ indexed over $ \mathbb N $, together with an action
  \[ \_\bullet\_ : P_m \times T_n^m \to P_n   \qquad \text{(note that we use the same notation as in \cref{def:algebraic-theory})}\]
  for all $ m, n $, such that for all $ t: P_l $, $ f: T_m^l $ and $ g: T_n^m $,
  \[
    t \bullet (x_{l, i})_i = t, \qand
    (t \bullet f) \bullet g = t \bullet (f_i \bullet g)_i.
  \]
\end{definition}

\noindent
A prime example of a $ T $-presheaf is $ T $ itself:
\begin{definition}[\coqident{AlgebraicTheories.Presheaves}{theory_presheaf}]\label{def:theory-presheaf}
  For an algebraic theory $ T $, its \iindex{theory presheaf} (also denoted $ T $) is the $ T $-presheaf given by the sets of $ T $. Its $ T $-action is the substitution operation of $ T $.
\end{definition}

\begin{definition}[\coqident{AlgebraicTheories.PresheafMorphisms}{presheaf_morphism}]
  For an algebraic theory $ T $, a \iindex{morphism $ f $ between $ T $-presheaves} $ P $ and $ Q $ is a sequence of functions $ f_n: P_n \to Q_n $ such that for all $ t: P_m $ and $ f: T_n^m $, $f_n(t \bullet f) = f_m(t) \bullet f$.
\end{definition}

\noindent
$T$-presheaves and their morphisms form the category of $ T $-presheaves \iindex{$ \Pshf T $} (\coqident{AlgebraicTheories.PresheafCategoryCore}{presheaf_cat}).
Because presheaf (iso)morphisms preserve $ \bullet $, we have, analogous to \cref{prop:algth-univ}:
\begin{lemma}[\coqident{AlgebraicTheories.PresheafCategory}{is_univalent_presheaf_cat}]
  $ \Pshf L $ is univalent.
\end{lemma}

\begin{definition}[\coqident{AlgebraicTheories.Examples.Plus1Presheaf}{plus_1_presheaf}]
  Given a $ T $-presheaf $ Q $, we can construct a presheaf $ A(Q) $ with $ A(Q)_n = Q_{n + 1} $ and, for $ q: A(Q)_m $ and $ f: T_n^m $, whose action is given by
  \[ q \bullet_{A(Q)} f = q \bullet_Q (\iota_{n, 1} (f_1), \dots, \iota_{n, 1} (f_m), x_{n + 1}). \]
  This is reminiscent of the $ \lambda $-theory axioms.
\end{definition}

\begin{lemma}[\coqident{AlgebraicTheories.RepresentationTheorem}{theory_presheaf_exponentiable}\coqident{AlgebraicTheories.PresheafCategory}{bin_products_presheaf_cat}]\label{exponential-presheaf}
The category of $ T $-presheaves has binary products: for presheaves $ P $ and $ Q $, we have $ (P \times Q)_n = P_n \times Q_n $ and $ (p, q) \bullet t = (p \bullet t, q \bullet t) $.
Furthermore, for all $ T $-presheaves $ Q $, $ A(Q) $ is the exponential object $ Q^T $.
\end{lemma}
\begin{corollary}[\coqident{AlgebraicTheories.RepresentationTheorem}{lambda_theory_to_reflexive_object}]
\label{cor:Hyland-reflexive}
The theory presheaf associated to a $\lambda$-theory is reflexive.
\end{corollary}
\begin{proof}
  Given a $ \lambda $-theory $L$, we have sequences of functions $ \lambda_n: A(L)_n \to L_n $ and $ \rho_n: L_n \to A(L)_n $.
  These commute with the $ L $-actions, so they constitute presheaf morphisms.
  Furthermore, by $ \beta $-equality, we have $\rho \circ \lambda = \id{A(L)} $.
\end{proof}

For ease of understanding, and to make Hyland's version of SRT particularly smooth, we defined presheaves in a very algebraic way. However, in category theory, a presheaf is commonly defined to be a contravariant, set-valued functor $ \op \C \to \SET $ on some category $ \C $. \cref{Lawvere-presheaf} justifies our definition by showing that $ T $-presheaves are equivalent to presheaves on some category:

\begin{definition}[\coqident{AlgebraicTheories.AlgebraicTheoryToLawvereTheory}{algebraic_theory_to_lawvere}]\label{lem:lawvere-clone}
  Let $ T $ be an algebraic theory. We construct the category $ \L $ with objects, morphisms, identity and composition for $ f: \L(l, m) $, $ g: \L(m, n) $ given by
  \[
    \L_0 = \{ 0, 1, 2, \dots \}, \quad
    \L(m, n) = T_m^n, \quad
    \id n = (x_i)_i \qand
    g \circ f = (g_i \bullet f)_i.
  \]
  We call $ \L $ the \iindex{Lawvere theory} associated to $ T $.
\end{definition}

Note that in $ \L $, the object $ n $ behaves as the $ n $-fold product $ 1^n $ with product projections $ \pi_{n, i} = x_{n, i}: \L(n, 1) $.

\begin{lemma}[\coqident{AlgebraicTheories.AlgebraicTheoryToLawvereTheory}{algebraic_presheaf_weq_lawere_presheaf}]\label{Lawvere-presheaf}
  Let $ T $ be an algebraic theory, and $ \L $ be its associated Lawvere theory as defined in \cref{lem:lawvere-clone}. The category $ \Pshf T $ of $ T $-presheaves is equivalent to the category $ P \L $ of presheaves (contravariant functors) on $ \L $.
\end{lemma}
\begin{proof}
  A $ T $-presheaf $ P $ corresponds to a $ \L $-presheaf $ Q $, where the sets $ P_n $ correspond to the action of $ Q $ on objects $ Q(n) $, and the $ P $-action $ P_m \times T_n^m \to P_n $ corresponds to the action of $ Q $ on morphisms $ \L(n, m) \times Q(m) \to Q(n) $.
\end{proof}

In the proof of the following lemma, we will make use of the so-called \iindex{Yoneda embedding} $ \yo : \C \hookrightarrow P \C $ of a category $ \C $ into its presheaf category. It sends an object $ X : \C $ to the functor $ Y \mapsto \C(Y, X) $. Note that it is fully faithful: it gives bijections on the morphisms $ \C(X, Y) \cong P \C(\yo(X), \yo(Y)) $. A well-known lemma in category theory is the \iindex{Yoneda lemma}, which shows that for $ X : \C $ and $ Y : P \C $, there is an equivalence $ P \C(\yo(X), Y) \simeq Y(X) $, which is natural in $ X $ and $ Y $.

\begin{lemma}[\coqident{AlgebraicTheories.RepresentationTheorem}{presheaf_lambda_theory_iso}]\label{lem:presheaf-Yoneda}
  For $ Q : \Pshf T $, we have bijections $ \Pshf T(T^n, Q) \cong Q_n $.
\end{lemma}
\begin{proof}
  Lemma \ref{Lawvere-presheaf} shows that $ T $-presheaves are equivalent to presheaves on the Lawvere theory $ \mathbf L $ associated to $ T $. Under this equivalence, $ \yo(n) $ corresponds to the power $ T^n = (T_m^n)_m $ of the theory presheaf, for all $ n : \L $. Then the Yoneda lemma gives a bijection $ \Pshf T(T^n, Q) \cong Q_n $. Explicitly, it sends $ f: \Pshf T(T^n, Q) $ to $ f_n(x_1, \dots, x_n) $ and $ q: Q_n $ to $ (t_i)_i \mapsto q \bullet t $.
\end{proof}

\noindent
Now we can give Hyland's proof of the representation theorem:
\begin{theorem}[\coqident{AlgebraicTheories.OriginalRepresentationTheorem}{endomorphism_theory_right_inverse}]\label{thm:Hyland}
  The function $ E $ of \cref{def:endomorphism-theory} has a right inverse $ H $, given by $ H(L) = (\Pshf L, L) $, where the last $ L : \Pshf L $ is the theory presheaf of \cref{def:theory-presheaf}.
\end{theorem}
\begin{proof}
  Let $ L $ be a $ \lambda $-theory. Recall from \cref{cor:Hyland-reflexive} that the theory presheaf $ L $ is a reflexive object with $ (L^L)_n = L_{n + 1} $, so $ E(L) $ has a $ \lambda $-theory structure.

  \cref{lem:presheaf-Yoneda} gives a sequence of bijections $ \varphi_n : \Pshf L(L^n, L) \cong L_n $ for all $ n $, sending $ f: \Pshf L(L^n, L) $ to $ f(x_1, \dots, x_n) $, and conversely sending $ s: L_n $ to $ ((t_1, \dots, t_n) \mapsto s \bullet (t_1, \dots, t_n)) $. It considers $ \lambda $-terms in $ n $ variables as $ n $-ary functions on the $ \lambda $-calculus. Therefore, $ \varphi $ preserves the $ x_i $, $ \bullet $, $ \rho $ and $ \lambda $, which makes it into an isomorphism of $ \lambda $-theories.
  Then, univalence of $\LamTh$ gives an identity $ E(S(L)) = L $.
\end{proof}

So Hyland shows that the representation theorem follows largely from the fact that the functions from $ L_n $ to itself can be represented by $ L_{n + 1} $, together with the Yoneda lemma for the Lawvere theory associated to $ L $ (\cref{lem:presheaf-Yoneda}).

Observe that the proof of \cref{thm:Hyland} does not require $ \beta $- or $ \eta $-equality, but if $ L $ has $ \beta $- or $ \eta $-equality, then we can immediately see (even without the isomorphism from the theorem) that the endomorphism theory also has this property.

  \section{The Karoubi Envelope}
\label{sec:karoubi}

In this section, we relate the construction of Hyland to that of Scott.
The relation is made precise by realizing that their categories are both strongly related to the \eemph{Karoubi envelope}.

Below, we characterize the Karoubi envelope via its defining universal property (\cref{classical-karoubi-envelope}).
In \cref{subsec:set-karoubi,subsec:univalent-karoubi}, we provide two implementations hereof.
Classically, these two constructions are equivalent.
In univalent foundations however, we see that the former provides a setcategory, whereas the latter provides a univalent category, and that one is the Rezk completion of the other (\cref{lemma:rezk-karoubi}).

Before stating the defining property of the Karoubi envelope, we first need the following definitions.
Let $ \C $ be a category and $ X, Y : \C $ objects. We will denote the type of section-retraction pairs of $ Y $ onto $ X $ with
\[ X \triangleleft Y = \sum_{r : \C(Y, X)} \sum_{s : \C(X, Y)} r \circ s = \id X. \]

Now, note that for $ (r, s) : X \triangleleft Y $, $s \circ r: \C(Y, Y) $ is an idempotent morphism, since $s \circ r \circ s \circ r = s \circ r $. We say that some idempotent morphism $ f: \C(X, X) $ \iindex{splits} if we can find some $ Y : \C $ and some $ (r, s) : X \triangleleft Y $ such that $ f = s \circ r $. If $ f $ does not split, a natural question to ask is whether we can find an embedding $ \iota : \C \hookrightarrow \D $ into some category $ \D $ such that the idempotent $ \iota(f): \D(\iota(X), \iota(X)) $ does split. This is one way to arrive at the \eemph{Karoubi envelope}. Its classical universal property is as follows:

\begin{definition}[\coqident{CategoryTheory.Categories.KaroubiEnvelope.Core}{karoubi_envelope}]\label{classical-karoubi-envelope}
  The \iindex{Karoubi envelope} of $ \C $ is a category $ \D $ such that
  \begin{enumerate}
    \item Every idempotent in $ \D $ splits;
    \item There is a fully faithful functor $ \iota: \C \to \D $;
    \item For every object $ Y : \D $, there merely exists an object $ X : \C $ and a retraction $ Y \triangleleft \iota(X) $.
      \setcounter{resume}{\value{enumi}}
  \end{enumerate}
\end{definition}

Classically, this determines the Karoubi envelope up to equivalence of categories.
However, as explained in \cref{subsec:univalence}, in univalent foundations, there are two notions of category: \eemph{setcategories} and \eemph{univalent categories}.
We hence add another condition:
\begin{enumerate}
  \setcounter{enumi}{\value{resume}}
  \item The setcategory Karoubi envelope (i.e.\ the Karoubi envelope in setcategory theory) should be a setcategory; the univalent Karoubi envelope should be a univalent category.
\end{enumerate}
In \cref{subsec:set-karoubi} (resp.\, \ref{subsec:univalent-karoubi}), we construct the setcategory (resp.\, univalent) Karoubi envelope.

\subsection{The Setcategory Construction}
\label{subsec:set-karoubi}

In this section, we construct a category $ \setkaroubi \C $ from a category $ \C $, and show that it is the setcategory Karoubi envelope if $ \C $ is a setcategory. The construction is the same as the one formalized in the 1lab \cite{onelab} (see also \cref{sec:related-work}).

\begin{definition}[\coqident{CategoryTheory.Categories.KaroubiEnvelope.SetKaroubi}{set_karoubi_cat}]\label{def:karoubi}
  We define the category $ \setkaroubi \C $. The objects of $ \setkaroubi \C $ are tuples $ (X, f) $ with $ X: \C $ and $ f: \C(X, X) $ such that $ f \circ f = f $. The morphisms between $ (X_1, f_1) $ and $ (X_2, f_2) $ are morphisms $ g: \C(X_1, X_2) $ such that $ f_2 \circ g \circ f_1 = g $. This can be summarized in the following diagram:
  \begin{equation}
    \begin{tikzcd}
      X_1
        \arrow["f_1"', loop, distance=2em, in=-150, out=150]
        \arrow[r, "g"] &
      X_2
        \arrow["f_2"', loop, distance=2em, in=30, out=-30]
    \end{tikzcd}
  \end{equation}
  The identity morphism on $ (X, f) $ is given by $ f $ and $ \setkaroubi \C $ inherits composition from $ \C $.
\end{definition}

\begin{theorem}[\coqident{CategoryTheory.Categories.KaroubiEnvelope.SetKaroubi}{set_karoubi}]
If $\C$ is a setcategory, then $ \setkaroubi C $ is a setcategory Karoubi envelope.
\end{theorem}
\begin{proof}
  \begin{enumerate}
	 \item Let $X = (Y,a) : \setkaroubi \C $ and $f : \setkaroubi \C(X, X)$ idempotent.
	 	Then $ f $ splits via $(f, f) : (Y, f) \triangleleft X$ since $f \circ f = f = \id{(Y,f)}$.
    \item Let $ \iota: \C \to \setkaroubi \C $ be the functor sending $ X: \C $ to $ (X, \id{X}) $ and $ f: \C(X, Y) $ to $ f $. It is fully faithful:
      \[ \setkaroubi \C((X, \id X), (Y, \id Y)) = \{ f: \C(X, Y) \mid \id Y \circ f \circ \id X = f \} \simeq \C(X, Y). \]
    \item Let $X = (Y,a) : \setkaroubi \C $ with $ Y: \C $ and $ a: Y \to Y $ idempotent. Then,
      $(a, a) : X \triangleleft \iota(Y)$,
      since $ a \circ a = a = \id{(Y,a)} $.
    \item If $ \C $ is a setcategory, the objects in $\setkaroubi{\C}$ form a set, since both the type of objects of $ \C $ and $\C(X,X)$ are sets, for any $ X : \C $.
    \qedhere
  \end{enumerate}
\end{proof}

\noindent
It is a general fact that completions induce monads:
\begin{remark}[\coqident{CategoryTheory.Categories.KaroubiEnvelope.SetKaroubi}{set_karoubi_monad}]
  The map $ \C \mapsto \setkaroubi \C $ constitutes a monad on the category of setcategories. The unit is given by the embedding functor $ \C \hookrightarrow \setkaroubi \C $. The multiplication $ \setkaroubi{\setkaroubi \C} \to \setkaroubi \C $ sends $ ((X, a), b) $ to $ (X, b) $, for $ X : \C $, $ a: \C(X, X) $ and $ b: \setkaroubi C((X, a), (X, a)) $.
\end{remark}

\begin{proposition}[\coqident{CategoryTheory.Categories.KaroubiEnvelope.SetKaroubi}{set_karoubi_univalence}]\label{rem:karoubi-univalent}
  If $ \setkaroubi \C $ is univalent, $ \C $ is univalent as well.
\end{proposition}
\begin{proof}
  The fully faithful embedding $ \iota : \C \hookrightarrow \setkaroubi \C $ induces equivalences $ (X \cong Y) \simeq (\iota(X) \cong \iota(Y)) $. We also have an equivalence $ (X = Y) \simeq (\iota(X) = \iota(Y)) $, because any identity $ X = Y $ also preserves the identity morphism on $ X $. Therefore, if $ \setkaroubi \C $ is univalent, we have a chain of equivalences
  $ (X = Y) \simeq (\iota(X) = \iota(Y)) \simeq (\iota(X) \cong \iota(Y)) \simeq (X \cong Y)$,
  which shows that $ \C $ is univalent as well.
\end{proof}

To show that the converse of \cref{rem:karoubi-univalent} does not hold, the next example considers the $1$-object category $\monoidcat{M}$ associated to a monoid $M$.
That is, $\monoidcat{M}$ has one object called, say, $\star$, morphisms $\monoidcat{M}(\star,\star) = M$, and the composition is given by the monoid multiplication.
\begin{example}\label{ex:karoubi-not-univalent}
  Consider the commutative monoid $ M $ consisting of the three matrices $ a = \left(\begin{smallmatrix}
      1 & 0\\0 & 1
    \end{smallmatrix}\right) $, $ b = \left(\begin{smallmatrix}
      1 & 0\\0 & 0
    \end{smallmatrix}\right) $ and $ c = \left(\begin{smallmatrix}
      -1 & 0\\0 & 0
    \end{smallmatrix}\right) $ under matrix multiplication.
    Then, $\monoidcat{M}$ is univalent since $a$ is the only isomorphism.
	However, $\setkaroubi{\monoidcat{M}}$ is \eemph{not} univalent; indeed, we have
	$\left( (\star,b) \cong (\star,b)\right) = \{b,c\} = \mathsf{2} \neq \mathsf{1} = \left( (\star,b) = (\star,b)\right)$.

\end{example}

\begin{example}[\coqident{AlgebraicTheories.LambdaTheoryToMonoid}{lambda_theory_to_monoid_cat_equality}]
\label{exa:R-eq-monoidcat-of-L1}
For $L$ a $\lambda$-theory, the category $\R$ (\cref{R}) is equal to $ \setkaroubi{\monoidcat{L_0}} $, where $ L_0 $ is the monoid $(\{ f : L_0 \mid \lambda(\rho(f)) = f \}, \circ, \lambda(x_1))$, consisting of terms with $ \eta $-equality.
\end{example}

\subsection{The Univalent Construction}
\label{subsec:univalent-karoubi}

In this section, we construct a category $ \univalentkaroubi \C $ from a category $ \C $, and show that it is the univalent Karoubi envelope of $ \C $, regardless of whether $ \C $ is univalent or not. We use the fully faithful Yoneda embedding $ \yo : \C \hookrightarrow P \C $.

\begin{definition}[\coqident{CategoryTheory.Categories.KaroubiEnvelope.UnivalentKaroubi}{univalent_karoubi_cat}]\label{def:karoubi''}
  We define the category $ \univalentkaroubi \C $ as the full subcategory of $ P \C $ consisting of objects $ F : P \C $ such that there merely exist an object $ X: \C $ and a retraction-section pair $ (r, s) : F \triangleleft \yo(X) $, summarized in the following diagram:
  \begin{equation}
    \begin{tikzcd}
      \mathit{\yo(X_1)} \arrow[r, "r_1", bend left, dashed] &
      F_1 \arrow[l, "s_1", bend left, dashed] \arrow[r, "g"] &
      F_2 \arrow[r, "s_2", bend left, dashed] &
      \mathit{\yo(X_2)} \arrow[l, "r_2", bend left, dashed]
    \end{tikzcd}
  \end{equation}
\end{definition}

\begin{theorem}[\coqident{CategoryTheory.Categories.KaroubiEnvelope.UnivalentKaroubi}{univalent_karoubi}]
  $ \R^U(C) $ is a univalent Karoubi envelope of $ C $.
\end{theorem}
\begin{proof}
  \begin{enumerate}
    \item \label{item:equalizer} Let $ f : \univalentkaroubi \C((Y, h_Y), (Y, h_Y)) $ be idempotent. Consider the equalizer $ (E, s) $ of $ f $ and $ \id Y $ in $ P \C $. Note that $ f \circ f = \id Y \circ f$, so by the universal property of the equalizer, there exists $ s : P \C(Y, E) $ such that $ f = s \circ r$.
      \begin{equation}
        \begin{tikzcd}
          E \arrow[r, "s"] & Y \arrow[r, shift left, "f"] \arrow[r, shift right, "\id Y"'] & Y\\
          Y \arrow[ru, "f"'] \arrow[u, dashed, "r"]
        \end{tikzcd}
      \end{equation}
      Note that $ s \circ (r \circ s) = f \circ s = s \circ \id E$, so again by the universal property of the equalizer, $r \circ s = \id E $, exhibiting $ E $ as a retract of $ Y $ in $ P \C $. Since we can compose retracts, and since $ (Y, h_Y) $ is in $ \univalentkaroubi \C $, we get some $ h_E $ such that $ (E, h_E) $ is in $ \univalentkaroubi \C $. Lastly, since the morphisms of $ \univalentkaroubi \C $ are borrowed directly from $ P \C $, we have $ f = s \circ r$ and $ f $ splits.
    \item The Yoneda embedding provides a functor $ \iota: \C \to \univalentkaroubi \C $, given by
      \[ \iota(X) = (\yo(X), \Vert (X, \id X, \id X) \Vert) \quad \text{and} \quad \iota(f) = \yo(f). \]
      Since the Yoneda embedding is fully faithful, $ \iota $ is fully faithful as well.
    \item For every $ (Y, h_Y) : \univalentkaroubi \C $, $ h_Y $ witnesses that there merely exist an object $ X : \C $ and a retraction $ (r, s) : Y \triangleleft \yo(X) $ in $ P\C $, and this gives a retraction $ (r, s) : (Y, h_Y) \triangleleft \iota(X) $.
    \item $\univalentkaroubi \C $ is univalent as a full subcategory of a univalent presheaf category.
    \qedhere
  \end{enumerate}
\end{proof}

\begin{remark}[\coqident{CategoryTheory.Categories.KaroubiEnvelope.UnivalentKaroubi}{idempotents_in_karoubi_envelope_split}]
Instead of \cref{item:equalizer} above, one can also show that idempotents split in $\univalentkaroubi{\C}$ by showing that idempotents split in $\SET$, and that, under certain conditions (including univalence of the target category), taking a functor category and a full subcategory preserves the splitting of idempotents.
\end{remark}

\begin{theorem}[\coqident{CategoryTheory.Categories.KaroubiEnvelope.RezkCompletion}{karoubi_rezk_completion}]
\label{lemma:rezk-karoubi}
  $ \univalentkaroubi \C $ is the Rezk completion of $ \setkaroubi \C $.
\end{theorem}
\begin{proof}
  $ \univalentkaroubi \C $ is univalent, and there is a fully faithful and essentially surjective functor $ \setkaroubi \C \to \univalentkaroubi \C $, sending $ (X, a) : \setkaroubi \C $ to the equalizer of $ \yo(\id X) $ and $ \yo(a) $.
\end{proof}

\subsection{The Relation Between Scott's and Hyland's Constructions} \label{subsec:scott-vs-hyland}
In this section, we will use the setcategory and univalent Karoubi envelopes to ``embed'' Scott's reflexive object into Hyland's reflexive object, and show that the correctness of Hyland's construction can be viewed as a consequence of the correctness of Scott's construction.

Recall that Scott and Hyland construct different right inverses to the function $E : \ReflOb \to \LamThType$.
Scott constructed the right inverse as $S(L) = (\R, U)$, whereas Hyland constructed $H(L) = (\Pshf{L}, L)$.
In this section, we show that $\R$ can be embedded into $\Pshf L$,
and that $U$ and $L$ coincide under this embedding.

Fix a $\lambda$-theory $L$.
Recall that $\R$ is equal to $\setkaroubi{\monoidcat{L_0}}$, where $L_0$ is the monoid $(L_0, \circ, \lambda(x_1))$; see \cref{exa:R-eq-monoidcat-of-L1}. We construct the embedding as a composition of the following functors:
\begin{equation}
  \begin{tikzcd}
    \R \arrow[r, "=", no head] & \setkaroubi{\monoidcat{L_0}} \arrow[rr, "\sim", no head] & & \setkaroubi{\monoidcat{L_1}} \arrow[d, two heads, hook] \\
    \Pshf L \arrow[r, "\sim", no head] & P \L \arrow[r, "\sim", no head] & P \monoidcat{L_1} & \univalentkaroubi{\monoidcat{L_1}} \arrow[l, hook]
  \end{tikzcd}
\end{equation}
where $L_1$ is the monoid $(L_1, \bullet, x_1)$. The following propositions construct two of the functors:
\begin{proposition}
[\coqident{AlgebraicTheories.LambdaTheoryToMonoid}{lambda_theory_to_monoid_karoubi_equiv}]
\label{lemma:R-equiv-KEML0}
The categories $\setkaroubi{\monoidcat{L_0}}$ and $\setkaroubi{\monoidcat{L_1}}$ are equivalent.
\end{proposition}
\begin{proof}
  The functions $ \rho: L_0 \to L_1 $ and $ \lambda : L_1 \to L_0 $ give an isomorphism of monoids $ L_0 \cong L_1 $, because elements of $ L_0 $ (resp. $ L_1 $) satisfy $ \eta $-equality (resp. $ \beta $-equality).
  The functoriality of $ \monoidcat - $ and $ \setkaroubi - $ turns this into an equivalence of categories $\setkaroubi{\monoidcat{L_0}} \simeq \setkaroubi{\monoidcat{L_1}}$.
\end{proof}

\begin{proposition}
[\coqident{AlgebraicTheories.PresheafEquivalence}{lawvere_theory_presheaf_equiv_monoid_presheaf}]
\label{lemma:PL-equiv-PL1}
Let $\D$ be a category with all colimits. Precomposition with the embedding $ \monoidcat{L_1} \hookrightarrow \L $ gives an equivalence between functor categories $[\L, \D]$ and $[\monoidcat{L_1}, \D]$.
\end{proposition}

Taking $ D = \op \SET $ gives an equivalence $P \L \simeq P(\monoidcat{L_1})$, because taking opposite categories preserves equivalences.
Together with \cref{lemma:R-equiv-KEML0}, \cref{lemma:rezk-karoubi}, the fact that $\univalentkaroubi{\monoidcat{L_1}}$ is a full subcategory of $P(\monoidcat{L_1})$ and \cref{Lawvere-presheaf},
this gives an embedding of $\R$ into $\Pshf{L}$:
\begin{theorem}[\coqident{AlgebraicTheories.RepresentationTheoremsRelation}{scott_to_hyland_fully_faithful}\coqident{AlgebraicTheories.RepresentationTheoremsRelation}{scott_to_hyland_U_to_L}]
\label{lemma:representation-thm-relation}
The composite of the functors
is fully faithful.
Furthermore, they send the reflexive object $ U : \R $ to the reflexive object $ L : \Pshf L $.
\end{theorem}

The embedding of \cref{lemma:representation-thm-relation} witnesses the passage from setcategories (the top row) to univalent categories (the lower row), and the vertical arrow in the diagram is precisely given by the Rezk completion. The theorem allows us to view $ \R $ as a full subcategory of $ \Pshf L $, where $ U : \R $ becomes $ L : \Pshf L $, which relates Hyland's proof to that of Scott:

Scott's and Hyland's constructions send a $ \lambda $-theory $ L $ to the reflexive objects $ S(L) = (\R, U) $ and $ H(L) = (\Pshf L, L) $ respectively. The endomorphism theory of a reflexive object $ (\C, X) $ is the $ \lambda $-theory $ E(L) $ with $ E(L)_n = \C(X^n, X) $. The result of this construction does not change when we pass to a larger category, as long as the powers of $ X $, the exponential $ X^X $ and the morphisms between them do not change. Therefore, one expects an isomorphism $ E(H(L)) \cong E(S(L)) $, which can be composed with the isomorphism $ E(S(L)) \cong L $ from Scott's version of SRT to get $ E(H(L)) \cong L $ for Hyland's version of SRT: if Scott's construction gives a section to the retraction $ E $, then Hyland's construction gives a section as well, even though the constructions themselves are quite different.

  \section{A Tactic for Propagating Substitutions}
\label{sec:tactic}

Now, as mentioned in \cref{lambda-calculus-operations}, any $ \lambda $-theory allows the operations $ \mathtt{var} $, $ \mathtt{app} $, $ \mathtt{abs} $ and $ \mathtt{subst} $ with the same interaction as for the pure $ \lambda $-calculus. Using these, we can define combinators like $ a \circ b $ or $ \langle a, b \rangle $ for $ a, b : L_0 $, used in the original proof of SRT. However the equalities about the interaction between $ \mathtt{subst} $ and the other operations are usually not definitional. Consider the following term (using concatenation for application):
$ \lambda x_5, x_5 (x_1 x_2 (x_4 x_3)) \bullet (x_1, x_2, x_3, x_1) : L_3$.
It is not hard to see that this results in $ \lambda x_4, x_4 (x_1 x_2 (x_1 x_3)) $. However, it takes several steps to rewrite this: moving the substitution past the $ \lambda $-abstraction, then into four applications, and lastly using five variable substitutions, resulting in a total of ten rewrites for a seemingly trivial term.
In other proofs, 40 or more of these rewrites need to be done, and since there are many such proofs, this quickly becomes tedious.
Therefore, we developed a tactic to perform these rewrites. A first version of this tactic was a variation of
\begin{lstlisting}
Ltac reduce_lambda := (rewrite subst_var + ... + rewrite beta_equality).
\end{lstlisting}
attempting to rewrite with some of the identities.
The statement \lstinline!repeat reduce_lambda! saved a lot of manual work, but it sometimes took a couple of seconds.

Therefore, we developed a new version of the tactic, called \lstinline!propagate_subst! (\coqident{AlgebraicTheories.Combinators}{propagate_subst}), written in \texttt{Ltac2} \cite{ltac2}.
It recursively traverses the $ \lambda $-term in the left-hand side of the goal, checking whether the term matches a form that can be rewritten into something else. It performs the possible rewrites, and also prints these rewrite statements which can replace it. For a small example, if the goal is about the interaction between the object $ U : \R $ and substitution,
\lstinline!U_term •  f = U_term!,
a call to \lstinline!unfold U_term; repeat reduce_lambda! takes ca.\ 100 ms, but \lstinline!propagate_subst ()! solves the goal in about 35 ms and prints
\begin{lstlisting}
refine '(subst_abs _ _ _ @ _).
refine '(maponpaths (λ x, (abs x)) (var_subst _ _ _) @ _).
refine '(maponpaths (λ x, (abs x)) (extend_tuple_inr _ _ _) @ _).
\end{lstlisting}
Replacing the call to \lstinline!propagate_subst! by these tactics results in the same rewrites, but these only take about five milliseconds.
Many equations in \coqfile{AlgebraicTheories.Combinators} were proved using the tactic. For example, the first 9 lines of \lstinline!subst_compose! take about 40 ms, whereas \lstinline!propagate_subst ()! and \lstinline!reduce_lambda! take about 330 ms and a second respectively. Apart from the performance gain at the expense of having longer proofs, another reason for replacing calls to \lstinline!propagate_subst! by the generated statements is the fact that the UniMath community aims to avoid extensive automation in the final proofs, to increase maintainability.

On top of the speedup, this tactic is modular: some of its parts are also tactics themselves, and can be reused for other tactics as well. For example, the \lstinline!traverse! tactic, which traverses a $ \lambda $-term in the goal and executes something for every subterm, is also used in another tactic we programmed that is called \lstinline!generate_refine!, which takes a pattern, and for every subterm that matches it, prints a statement such as 
\begin{lstlisting}
 refine '(maponpaths (λ x, ... x ...) _ @ _).
\end{lstlisting}
which can be used to quickly generate statements that very precisely rewrite one subterm.

The \lstinline!propagate_subst! tactic is also extensible: the patterns for both the subterm traversal and the rewrites are kept in a list, which can be extended when new combinators are defined. For example, at the point where the tactics are defined, the traversal only works for the constructors $ \mathtt{var} $, $ \mathtt{app} $, $ \mathtt{abs} $ and $ \mathtt{subst} $, and the rewrites only work for the interactions between those. Using this, composition $ a \circ b $ is defined, and so a pattern to branch into $ a $ and $ b $ is added to the traversals, and a rewrite with
$(a \circ b) \bullet t = (a \bullet t) \circ (b \bullet t)$
is added. Progressing through the file, the same is done for combinators like the pair $ (a, b) $, the projection $ \pi_1 $ (including a rewrite $ \pi_1 \langle a, b \rangle = a $), $ \mathtt{curry} $ and $ \mathtt{n\_tuple} $ (consisting of nested pairs).

We generalized \lstinline!propagate_subst! to a tactic for simplification with arbitrary sets of identities (\coqident{Tactics.Simplify}{simplify}), and applied this to formulas in a hyperdoctrine (\coqident{CategoryTheory.Hyperdoctrines.FirstOrderHyperdoctrine}{hypersimplify0}), a formalization related to \cite{niels-kobe-Rezk-abstract}.

  \section{Related Work}
\label{sec:related-work}

We first discuss related work in the area of \textbf{formalization of denotational semantics} of lambda calculi, with particular focus on fixpoints.
%
%

Perhaps most relevant to our work, Benton, Kennedy, and Varming \cite{domain-theory-coq} formalize, in Rocq, a constructive notion of $\omega$-cpos and the construction, via limits, of solutions to recursive domain equations in the category of $\omega$-cpos.
This involves constructing fixpoints on the level of types.
Similarly, Dockins \cite{DBLP:conf/itp/Dockins14} formalizes effective domain theory in Rocq.
Those works are thus complementary to ours: the authors work exclusively in the category of certain domains (e.g., $\omega$-cpos), whereas we study general reflexive objects in suitable categories: their specific $\omega$-cpos could be fed into our \eemph{endomorphism theory} function to obtain specific lambda calculi.

De Jong \cite{DBLP:journals/mscs/Jong21} models PCF in domains, formalizing the Scott model in particular.
In \cite{DBLP:conf/csl/JongE21}, De Jong and Escardó construct Scott's $D_\infty$ in univalent foundations, in Agda.

Møgelberg \cite{DBLP:conf/icalp/Mogelberg06} develops a language with fixed point combinators as a metalanguage to reason about domains.
In another line of work, \emph{guarded} type theory (see, e.g., \cite{DBLP:journals/corr/abs-1208-3596,DBLP:journals/mscs/MogelbergP19}) is being developed as a meta-language for reasoning about recursive domain equations.

Less closely related to our work on the technical level are other efforts formalizing the syntax and semantics of programming languages specifically in UniMath.
In particular, Ahrens, Lumsdaine, and Voevodsky \cite{DBLP:journals/lmcs/AhrensLV18} compare, in UniMath, different notions of model for dependent type theories.
Van der Weide \cite{DBLP:journals/corr/abs-2411-06636} constructs an equivalence of bicategories between univalent locally cartesian closed categories and univalent categories with suitable type formers, and various extensions of that equivalence to additional structure.

Finally, Altenkirch, Kaposi, Šinkarovs, and Végh \cite{altenkirch_et_al:LIPIcs.FSCD.2023.24} construct, in Cubical Agda, an equivalence between the simply-typed lambda calculus and combinatory logic.
Both theories are given as Generalized Algebraic Theories.

We mention some work on formalization of \textbf{initial algebra semantics} for languages including the untyped lambda calculus.
Hirschowitz and Maggesi \cite[Thm.~3]{hirscho-maggesi} formalize the untyped lambda calculus with $\beta$- and $\eta$-equality.
Ahrens \cite{UntypedRelativeMonads16} proves an initial semantics result including, as a guiding instance, the lambda calculus with $\beta$- and $\eta$-\emph{reduction} (as opposed to $\beta$- and $\eta$-\emph{equality}), formalized in Rocq.
Ahrens, Hirschowitz, Lafont, and Maggesi \cite{2Signatures19,PresentableSignatures18} prove  initial semantics for abstract syntax with equations, and formalize their results in UniMath.

Next, we discuss related work in the area of \textbf{formalization of (univalent) category theory}.
The formalization of univalent category theory started with \cite{univalent-categories}, and was continued in many different works, e.g., \cite{displayed-categories,DBLP:conf/cpp/AhrensMWW24,DBLP:journals/mscs/AhrensFMVW21,DBLP:conf/cpp/WeideRAN24}, by various authors.
We are basing our formalization on the library of univalent category theory built to accompany these formalizations.

There are several other libraries on univalent category theory.
The HoTT library \cite{DBLP:conf/cpp/BauerGLSSS17} focuses on ``wild'' (higher) categories, that is, categories that omit truncation levels for types of cells.
For our purpose, it is important to work with ``true'' categories, that is, things that correspond to categories in Voevodsky's simplicial set model of univalent foundations \cite{kapulkin2021simplicial}.
There does not seem to be a formalization of the Karoubi envelope in the HoTT library.
The 1lab \cite{onelab} is another library of univalent category theory. It features a formalization of the Karoubi envelope of a \emph{pre}category at \url{https://1lab.dev/Cat.Instances.Karoubi.html}.
Here, a ``precategory'' is essentially a category without the assumption that it is either a setcategory or a univalent category.
The construction implemented there is the one described in \cref{subsec:set-karoubi}; as discussed in \cref{ex:karoubi-not-univalent}, it does not necessarily yield a univalent category.

There are many other libraries of formalized category theory; for space reasons, we only point to papers listing and comparing some of them.
Gross, Chlipala, and Spivak \cite{DBLP:conf/itp/GrossCS14} provide a comparison of different libraries that, at publication time, was quite comprehensive.
Hu and Carette \cite{DBLP:conf/cpp/HuC21} provide another comparison of different libraries.

Finally, Shulman \cite{DBLP:journals/corr/Shulman15} studies the question whether idempotent functions between types in MLTT split, with a formalization in Rocq.
Semantically, that work is concerned with morphisms between $\infty$-groupoids, not with morphisms in a 1-category as in our setting.

  \section{Conclusion}

We have presented a formalization of SRT in univalent foundations.
We see two benefits to working in univalent foundations:
\begin{enumerate}
\item We can express on the level of \emph{types} a meaningful relation between $\lambda$-theories and reflexive objects, by stating that the functions of Diagram \eqref{diag:intro-diag} form a section-retraction pair. Proving this is possible because, starting with a given $\lambda$-theory, going to reflexive objects and back to $\lambda$-theories yields an isomorphic --- and hence identical --- $\lambda$-theory. This uses crucially that the category of $\lambda$-theories is univalent, that is, that isomorphism of $\lambda$-theories is the same as their identities --- a consequence of the univalence axiom.
\item The differences between Scott's and Hyland's constructions are clarified in univalent foundations: Scott's is adequate for setcategories, Hyland's for univalent categories.
\end{enumerate}
We have furthermore programmed a tactic to manipulate lambda terms, combining the convenience of automation with the good performance of precise rewriting.

There are some questions we have left open.
In particular, one could ask how Diagram \eqref{diag:intro-diag} lifts to the level of \emph{categories}.
In that case, the left-hand side clearly forms a univalent 1-category (\cref{prop:lamth-univalence}).
The status of the right-hand side is less clear: it could form a (univalent) 1-category whose objects are certain setcategories with an object therein, or a bicategory whose objects are certain univalent categories.

  \newpage
  \bibliography{citations}

\end{document}